\documentclass[twocolumn,showpacs,preprintnumbers,amsmath,amssymb,
 aps,
 prb,
 lengthcheck,
]{revtex4}
\usepackage{color}
\usepackage{graphicx}
\usepackage{dcolumn}
\usepackage{bm}
\usepackage{hyperref}
\usepackage{makeidx}
\makeindex

\def\>{\right\rangle}
\def\<{\left\langle}
\def\be{\begin{equation}}
\def\ee{\end{equation}}
\def\ba{\begin{array}{lll}}
\def\ea{\end{array}}
\def\f{\frac}

\def\beq{\begin{eqnarray}}
\def\eeq{\end{eqnarray}}
\begin{document}

\title{Finite frequency noise for edge states at filling factor $\nu=2/5$}
\author{D Ferraro$^{1,2,3}$ A Braggio$^{3}$, N Magnoli$^{1,2}$ and M Sassetti$^{1,3}$}
\address{$^1$ Dipartimento di Fisica, Universit\`a di Genova, Via Dodecaneso 33, 16146 Genova, Italy}
\address{$^2$ INFN, Sezione di Genova, Via Dodecaneso 33, 16146 Genova, Italy}
\address{$^3$ SPIN-CNR, Via Dodecaneso 33, 16146 Genova, Italy}
\begin{abstract}
  We investigate the properties of the finite frequency noise in a quantum point contact geometry for the fractional 
quantum Hall state at filling factor $\nu=2/5$.  The results are obtained in the framework of the Wen's hierarchical model. 
  We show that the peak structure of the colored noise allows to discriminate among different possible excitations 
involved in the tunneling. In particular, optimal values of voltage and temperature are found in order to enhance the 
visibility of the peak associated with the tunneling of a 2-agglomerate, namely an excitation with charge double of the 
fundamental one associated to the single quasiparticle.
\end{abstract}

\maketitle

\section{Introduction}
 The Fractional Quantum Hall Effect (FQHE) \cite{DasSarma97} is one of the most remarkable examples of strongly
correlated electron system. Since its discovery it has been subject on an intense study both from the theoretical and the experimental point of view, with the aim of providing an unified picture of the plethora of different observed values of filling factor $\nu$. \cite{Tsui99} A suitable theoretical framework for the description of these states is provided by the theory of edge states \cite{Wen95}. For the Laughlin series \cite{Laughlin83} at $\nu=1/(2n+1)$ ($n\in \mathbb{Z}$) a
 chiral Luttinger Liquid theory described in terms of a single bosonic mode was
 proposed. \cite{Wen90}  For the more general Jain series \cite{Jain89} with  $\nu=p/(2np+1)$ ($p\in \mathbb{Z}$) a possible extension has been proposed by X. G. Wen  \cite{Wen92} considering $|p|-1$ additional hierarchical fields, propagating with finite velocity along the edge.
 
 Noise experiments in the Quantum Point Contact (QPC) geometry \cite{Chang03} have been
 crucial to demonstrate the existence of the peculiar fractionally charged excitations predicted by the theoretical models. \cite{Laughlin83}  
 In particular it was proved that, for the Jain sequence, the elementary quasiparticle (qp)
 charge is given by $e^*=e/(2np+1)$. \cite{dePicciotto97, Saminadayar97, Reznikov99} More recently some intriguing transport experiments have been performed on a QPC at very low temperature and extremely weak backscattering for various filling factors showing a change in the power-law of the backscattering current as a function of temperature and an enhancing of the effective tunneling charge, as extracted from the noise. \cite{Chung03, Bid09, Dolev10} In order to explain both these unexpected behaviors we observed that propagating neutral modes could crucially modify the scaling dimensions of operators leading to a crossover in the relevance of the tunneling process according to measurements.\cite{Ferraro08, Ferraro10a, Ferraro10b, Ferraro10c, Carrega11}

The symmetrized noise at finite frequency \cite{Rogovin74} represents an important tool to characterize the predicted crossover phenomenology, complementary and alternative to current and zero frequency noise. From the experimental point of view, measurements of colored noise
have been carried out for a QPC in a 2D electron gas at zero magnetic field
\cite{Zakka07}.  Nevertheless, great efforts are devoted to extend the
observations to more interesting cases including the FQH regime. From a theoretical point of view, the colored noise was investigated for the Laughlin
sequence \cite{Chamon95, Chamon96}  and also for more exotic states like $\nu=5/2$. \cite{Bena06, Carrega12}  In these cases, much of the relevant physics occurs at frequencies close to Josephson frequencies multiple of
$\omega_{0}=e^{*}V/\hbar$, ($e^{*}$ the charge of the elementary excitation) that is in the range of GHz for external
bias $V$ in the range of $\mu$V.  These values,
although high, should be  experimentally observable. 

In this paper we will discriminate among the different tunneling charges
involved in the transport, being their contributions resolved at different
frequencies.  We will focus on the edge state at $\nu=2/5$ where only two bosonic fields, 
one charged and one neutral, are involved. \cite{Wen95} We will demonstrate  the necessity to consider voltages larger than the 
neutral modes cut-off frequency, in order to efficiently detect the presence of the 2-agglomerate. 

The paper is divided as follows. In Sec. \ref{sec: wen_jain} we recall the Wen's model 
for the description of the FQH state at $\nu=2/5$.
In Sec. \ref{Transport} we provide the expression for the backscattering current and noise in terms of the tunneling rates. 
In Sec. \ref{scaling_sec} we analyze the scaling behaviors  for various energy regimes. 
In Sec. \ref{results} we present the finite frequency noise peak structure for $\nu=2/5$ as a function of the bias voltage in order 
to determine the optimal values of the physical parameters to observe the $2$-agglomerate peak. Sec. \ref{conclusion} is devoted 
to the conclusions. 

\section{The model}\label{sec: wen_jain}
We consider infinite edge states of an Hall bar in the Jain sequence at filling factor $\nu=p/(2np+1)$. \cite{Jain89} We focus on the case $\nu=2/5$ ($n=1$ and $p=2$), described by the Lagrangian density \cite{Wen95, Ferraro10b} ($\hbar=1$)
\be
\mathcal{L}=\f{5}{8\pi} \partial_{x}\varphi^{\rm{c}}\left(
\partial_{t}-v_{\rm{c}}\partial_{x}\right)\varphi^{\rm{c}}+\f{1}{8\pi} \partial_{x}\varphi^{\rm{n}}\left(\partial_{t}-v_{\rm{n}}
\partial_{x}\right)\varphi^{\rm{n}},
\label{action}
\ee
with $v_{\rm{c}}$  the propagation velocity of the charged bosonic mode $\varphi^{\rm{c}}$ and  $v_{\rm{n}}$ the one of
the neutral bosonic mode $\varphi^{\rm{n}}$. The electron number density of the edge depends on the charged 
field only, via the relation $\rho(x)=\partial_{x}\varphi^{\rm{c}}(x)/2\pi$. Due to this fact the charge mode 
velocity is affected by Coulomb interactions, leading to the reasonable 
assumption $v_{\rm{c}}\gg v_{\rm{n}}$.\cite{Levkinskyi08, Levkinskyi09, Ferraro08} From (\ref{action}) one can
 note that the neutral mode co-propagates with respect to the charged one as typical of the states  
 with $p>0$. \cite{Wen95} The commutators of  the bosonic fields are $[\varphi^{\rm{c/n}}(x),\varphi^{\rm{c/n}}(y)]=i\pi \nu_{\rm{c/n}} 
\mathrm{sgn}(x-y)$ with $\nu_{\rm{c}}=2/5$ and $\nu_{\rm{n}}=2$. 

We consider a generic $m$-agglomerate annihilation operator with charge  $e^{*}_m =  (e m \nu_c )/2  = m e^{*}$ with  $e^{*} =  e /5$ the charge
 of the elementary excitation, namely the single-quasiparticle (qp) ($e$ the electron charge). It can be written as
\be
\label{op_wen_5}
\Psi^{(m,j)} (x,t) = \frac{\mathcal{F}^{(m,j)}}{\sqrt{2 \pi a}} e^{i \left( \frac{m}{2} \varphi^{\mathrm{c}} (x,t) + \frac{j}{2} \varphi^{\mathrm{n}} (x,t)
 \right)}
\ee
where $a$ is a finite length cut-off, $m \in \mathbb{N}$ is related to the charge of the excitation and the additional quantum number $j \in \mathbb{Z}$ 
plays a role analogous to the isospin and is necessary in order to provide the proper fractional statistics. \cite{Ferraro10c} Note that for a given quasiparticle excitation the integer $m$ and $j$ must have the same parity. The Klein factors $\mathcal{F}^{(m,j)}$ are ladder operators necessary in order to change the particle numbers of the $m$-agglomerate and to provide the proper statistical properties between excitations. \cite{Ferraro10a, Ferraro10c} From the long time limit of the
two-point imaginary time Green's function $\mathcal{G}^{(m,j)}(\tau)=\<T_{\tau}\Psi^{(m, j)} (0,\tau){\Psi^{(m,j)}}^{\dagger}(0,0)\>$ calculated at $x=0$ and
at zero temperature \cite{Kane92}, one can extract the scaling dimension of the $m$-agglomerate

\be
\Delta^{(m, j)}=\frac{1}{2} \left[\nu_{\mathrm{c}} \left(\frac{m}{2}\right)^{2}+\nu_{\mathrm{n}} \left(\frac{j}{2}\right)^{2}\right].
\label{scaling}
\ee

For energies higher than the typical neutral mode bandwidth $\omega_{\mathrm{n}}=v_{\mathrm{n}}/a$ the neutral mode contribution to the dynamics saturates and one has the effective scaling dimension \cite{Ferraro08, Ferraro10a}
\be
\tilde{\Delta}^{(m, j)}=\frac{1}{2} \nu_{\mathrm{c}}\left( \frac{m}{2}\right)^{2}
\label{scaling_eff}
\ee  
that only depends on the charged sector of the theory. From now on the correspondent charged mode bandwidth $\omega_{\mathrm{c}}=v_{\mathrm{c}}/a$ is assumed as the greatest energy scale of the system.

Note that the above scaling dimension could be strongly affected by interaction with the external environment \cite{Rosenow02, Papa04, Mandal02, Yang03, Cuniberti97, Ferraro08, Carrega11, Braggio12}, nevertheless in this paper, for sake of simplicity, we only consider the standard unrenormalized case.

\section{Transport properties} \label{Transport}
The tunneling through the QPC at $x=0$ of a generic $m$-agglomerate between the $R$ and $L$ edges of the
Hall bar, is described by the Hamiltonian
$H^{(m)}_{T}=\textbf{t}_{m}\Psi^{(m)}_R(0){\Psi_L^{(m)}}^{\dagger}(0)+{\rm
 h.c.}$, where $\Psi^{(m)}_R$  ($\Psi^{(m)}_L$) is the $m$-agglomerate annihilation operator. From equation (\ref{scaling}) it is easy to note that the more relevant single-qps ($m=1$) are the ones with $j=\pm 1$ ($\Delta^{(1,\pm1)}=3/10$), while among all the $2$-agglomerate ($m=2$) we consider the one with $j=0$ ($\Delta^{(2,0)}=1/5$). All the other operators with the same charge (same $m$) and different $j$ are less relevant in the renormalization group sense and give a negligible contribution to the transport properties with respect to the considered one. \cite{Ferraro08} From now on, for notational convenience, we will omit the index $j$ where not necessary. 

The tunneling amplitudes $\textbf{t}_{m}$ depend, in general, on the geometry of the constriction, 
and can be energy dependent. \cite{Chang03} 
In the following they  will be assumed as $m$-dependent constants in order to make the discussion 
clearer and to restrict the number of free parameters of the theory. The total 
tunneling Hamiltonian will consist of the sum over all possible $m$-agglomerate $H_T=\sum_{m}H_T^{(m)}$.

In the following we will focus on the weak tunneling regime. At lowest order in the tunneling Hamiltonian  the
backscattering current of the $m$-agglomerate $I^{(m)}_{B}$ and the finite frequency symmetrized noise $S_{B}^{(m)}(\omega)$ 
can be written in terms of the tunneling rates \cite{Rogovin74}. By using the detailed balance 
relation the current is \cite{Ferraro10a} ($k_{B}=1$) 
\be 
I^{(m)}_{B}(\omega_0)=m e^{*}\left(1-e^{-m\omega_0/T}\right)\mathbf{\Gamma}^{(m)}( \omega_0)\,,
 \label{current_bal} 
 \ee 
with tunneling rate
\be
\mathbf{\Gamma}^{(m)}(E)=|\textbf{t}_{m}|^{2}
\!\int^{+\infty}_{-\infty} \!\!\! dt e^{im Et } G^{<}_{m,R}(0,-t) G^{>}_{m, L}(0,t)\,.
\label{tunn_rate}
\ee 
Here, $\omega_0 = e^* V$ is the Josephson frequency associated to the single-qp with $V$ the bias. 
The correlators $G_{m,l}^{>}(0,t)=\langle \Psi^{(m)}_{l}(0,t)
\Psi^{(m)\dagger}_{l}(0,0)\rangle=(G^{<}_{m,l}(0,t))^*$ are the
two point Green's functions of the $m$-agglomerate operators on the edge $l=R,L$. 

The noise spectral density is
\be
S_{B}^{(m)}(\omega)=\int_{-\infty}^{+\infty} dt \, e^{i \omega t}S_{B}^{(m)} (t)\,, 
  \ee 
  where  
  \be S_{B}^{(m)} (t ) = \left[ \langle \delta I^{(m)}_B (t) \delta I^{(m)}_B (0) \rangle +
  \langle \delta I^{(m)}_B (0) \delta I^{(m)}_B (t) \rangle \right]\,,
\label{symm_noise}
\ee

with $\delta I^{(m)}_B$ the fluctuations of the current with respect to the average. 

Recalling the definition of the tunneling rate in (\ref{tunn_rate}) one has 
\be
S_{B}^{(m)}(\omega)=(me^{*})^{2}\sum_{\varepsilon, \eta=\pm} \left[\mathbf{\Gamma}^{(m)}(\varepsilon\omega/m+ \eta \omega_0 )\right]\,,
\label{Noise_freq_rate}
\ee 
or, equivalently, in terms of the backscattering current
 \beq
S_{B}^{(m)}(\omega)&=&me^{*}\sum_{\varepsilon=\pm}\coth{\left(\frac{\varepsilon\omega+
      m \omega_0 }{2 T}\right)} \nonumber \\
      &\times& I_B ^{(m)} (\varepsilon\omega/m+\omega_0 )\,.
\label{Noise_freq_curr}
\eeq 
The above result is consistent with the non-equilibrium
fluctuation-dissipation theorem. \cite{Rogovin74, Dolcini05}  Note that from (\ref{Noise_freq_rate}) one can easily 
restore the well known result
for the zero frequency limit. \cite{Ferraro10a, Martin04, Safi01}

At  lowest perturbative order the total noise will be 
\be
S_{B}(\omega)=\sum_{m}S^{(m)}_{B}(\omega)\,,
\label{totalnoise}
\ee
being the contributions of the different $m$-agglomerate independent.  Note that, analogously, the total  backscattering current  is given by $ I_{B} =\sum_{m} I^{(m)}_{B} $. The simple relations in (\ref{Noise_freq_curr}) - (\ref{totalnoise}) are due to the Poissonian statistics of the  tunneling processes at lowest order in $|\textbf{t}_m|^2$ and to the independency of the sources of noise.

\section{Scaling behavior}\label{scaling_sec}
Let us now focus on the evaluation of the tunneling rate in (\ref{tunn_rate}),
starting from the zero temperature limit.  Here, the bosonic Green's functions
are \cite{Braggio01, Cavaliere04, Cavaliere04b} 
\beq
\langle \varphi^{s}(0,t) \varphi^{s}(0,0) \rangle&=-\nu_{s} \ln{
  (1+i \omega_{s} t)},
\label{GFbosonic} 
\eeq 

with $s=\mathrm{c}, \mathrm{n}$, leading to the tunneling rate \cite{Ferraro08}
\beq
\mathbf{\Gamma}^{(m)}(E)&& =
\frac{ |\textbf{t}_m|^2}{2 \pi a^2} \left(\frac{m E}{\omega_{\mathrm{c}}}\right)^{\alpha}\!\left( \frac{mE}{\omega_{\mathrm{n}}}\right)^{
\delta} \!e^{-\frac{m E}{\omega_{\mathrm{c}}}} \frac{\left(m E\right)^{-1}}{\Gamma (\alpha+\delta)} \nonumber \\
&& _1 F_1\! \left[\delta; \alpha+\delta; \left(\frac{m E}{\omega_{\mathrm{c}}} -\frac{m E}{\omega_{\mathrm{n}}}\right) 
\right]\Theta(m E)\,.
\label{risoluzione_due_modi}
\eeq 
Here,$\ _1F_1[a;b;z]$ is the Kummer confluent hypergeometric
function \cite{gradshteyn94}, $\alpha=\nu_{\mathrm{c}}m^2/2$ and $\delta =  \nu_{\mathrm{n}} j^2/2$ the charged and the neutral
exponent respectively.

The rate  shows two different regimes.
At low energies ($E\ll\omega_{\rm n}$) 
the rate scales as 
\be
\mathbf{\Gamma}^{(m)}(E) \approx E^{4 \Delta^{(m,j)}-1} 
\ee
receiving contributions from both the charged and the neutral modes. In the
same limit, for frequencies close to the Josephson resonance $\omega\to
m\omega_0$, one has (cf. (\ref{Noise_freq_rate}))
\be
S_{B}^{(m)}(\omega\rightarrow m \omega_0)\approx (\omega- m \omega_0
)^{4 \Delta^{(m,j)}-1}\,.
\label{noiseJoseph}
\ee

As stated before, for the single-qp the scaling dimension is given by $\Delta^{(1,1)}=3/10$, while for the next excitation, the $2$-agglomerate, the scaling are driven by the charged mode contribution only in such a way that $\Delta^{(2,0)}=1/5$ (cf. (\ref{scaling})).
These behaviors indicate the $2$-agglomerate as the dominant excitation in the current.\cite{Ferraro08} All other excitations, with higher charges and different isospin, have higher scaling dimensions and can be safely neglected.

Note that the peculiar values of the scaling dimensions imply a divergent power-law behavior of the total
noise in correspondence of the $2$-agglomerate Josephson frequencies ($2\omega_{0}$) and a dip for the single-qp ($\omega_{0}$). 

For applied voltages higher than the neutral mode cut-off ($\omega_0>\omega_{\mathrm{n}}$) the neutral modes saturate leading to a lower effective dimension $\tilde{\Delta}^{(1,1)}=1/20$ for the single-qp (cf. (\ref{scaling_eff})). On the other hands, the 2-agglomerate scaling is unaffected. In this case one has a two peak structure in correspondence of the Josephson frequencies of the two excitations ($\omega_{0}$ and $2\omega_{0}$ respectively). 

At finite temperature the above behaviors will be smoothened with more remarkable changes near to the Josephson 
resonances $\omega=m\omega_0$ for $T>|\omega-m\omega_0|$.
To quantitatively determine the temperature influence on the  noise one has to consider  
the finite temperature expressions for the Green's functions in (\ref{GFbosonic})  
for the bosonic fields. \cite{Ferraro08, Ferraro10a} 
  
\be
\langle \varphi^{s}(0,t) \varphi^{s}(0,0) 
\rangle=\nu_{s} \ln{\left[\frac{|\Gamma\left(1+T/\omega_{s}-i tT\right)|^{2}}{\Gamma^{2}\left(1+T/\omega_{s}\right)
 (1+i \omega_{s} t)}\right]} 
 \ee
 with $s=\mathrm{c}, \mathrm{n}$.
 
The tunneling rate can be still analytically evaluated
for temperatures lower than the bandwidths, namely
$T\ll\omega_{\mathrm{n}}, \omega_{\mathrm{c}}$, leading to 
\beq
\mathbf{\Gamma}^{(m)}(E) &=&\frac{|t_m|^2}{(2 \pi a)^2}\frac{ (2 \pi) ^{\alpha+ \delta}}{\omega_{\mathrm{c}}^{\alpha} 
\omega_{\mathrm{n}}^{\delta}}T^{\alpha+\delta-1}e^{\frac{mE}{2T}} \nonumber \\
 &\times& B\left(\alpha+\delta-i\frac{mE}{2 \pi T}; \alpha+\delta+i\frac{mE}{2 \pi T}\right) \nonumber
 \\
 \eeq
being $B(a; b)$ the Euler beta function \cite{gradshteyn94} ($\alpha=\nu_{\mathrm{c}}m^2/2$ and $\delta =  \nu_{\mathrm{n}} j^2/2$). At higher temperatures the rate have 
to be evaluated numerically.\cite{Ferraro10a}

Close to the Josephson frequencies, the noise in (\ref{Noise_freq_curr}) reduces to
\be
S_{B}^{(m)}(\omega\rightarrow m \omega_0 )\approx TG_{B}^{(m)}(T)
\label{scaling_peaks}
\ee 
written in terms of the $m$-agglomerate linear conductance 
\be
G_{B}^{(m)}(T)=(m e^{*})^{2} \frac{\mathbf{\Gamma}^{(m)}(0)}{T}.
\ee
The power-law behavior for the height of the peaks as a function of the temperature is therefore given by
$S_{B}^{(m)}(\omega\rightarrow m \omega_0 )\approx T^{\alpha+\delta-1}$ for $T\ll \omega_{\mathrm{n}}$ and 
$S_{B}^{(m)}(\omega\rightarrow m \omega_0 )\approx T^{\alpha-1}$ for $T\gg \omega_{\mathrm{n}}$.

The visibility of the $2$-agglomerate peak is guaranteed in this regime for $S_{B}^{(1)}(2\omega_0)<S_{B}^{(2)}(2\omega_0)$. 

 \begin{figure}
   \centering
 \includegraphics[width=0.5 \textwidth]{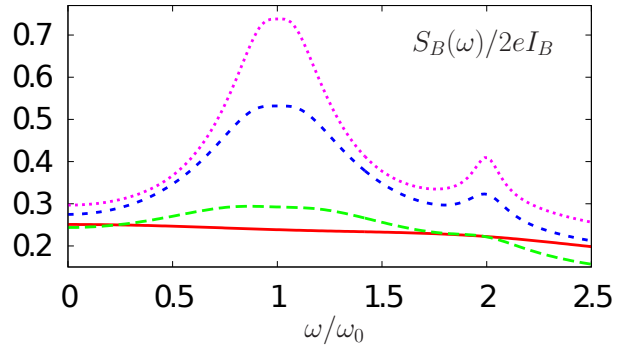}
 \caption{Spectral noise $S_B (\omega )/2 e I_{B}$ as a function of the normalized frequency $\omega/\omega_{0}$. Different values of the bias are given by:
 $\omega _0 = 30$ mK (solid red); $\omega _0 = 100 $ mK (dashed green); $\omega _0 = 200 $ mK (short-dashed blue); $\omega _0 = 300 $ mK (dotted magenta). Other parameters are: $\omega _{\mathrm{c}}=5$ K, $\omega _{\mathrm{n}} = 50 $ mK, $|t_{2}|/|t_{1}|=1$, $T = 10 $ mK.}
  \label{Fig1}
 \end{figure}

\section{Results}
\label{results}
As stated before, for the composite edge states belonging to the Jain sequence, important information about 
the carrier charges involved in the tunneling can be extracted from the knowledge of the total noise at finite frequency in (\ref{totalnoise}). Each $m$-agglomerate contribution leads to a resonance at $\omega=m \omega_{0}$. This enable us to 
resolve them and to obtain indications of their power-law behavior. Depending on their scaling dimensions, they show peaks (for $\Delta<1/4$), dips (for $1/4<\Delta<1/2$) or a monotonic increasing as a function of frequency (for $\Delta>1/2$). 
Note that, for notational convenience, we generically indicated with $\Delta$ both the low energy 
scaling dimension in (\ref{scaling}) and the effective scaling above the neutral 
mode bandwidth in (\ref{scaling_eff}) depending on the considered regime. In figure \ref{Fig1} we show the behavior of the spectral noise for the 
considered $\nu=2/5$ case at low temperature ($T=10$ mK). 

The ratio $S_B (\omega )/2 e I_{B}$ is plotted  in order 
to recover the proper value of the Fano factor in the zero frequency limit. \cite{Ferraro10a, Carrega12} To further 
simplify the description, we  only focus on the first two contributions to the noise, namely single-qp and $2$-agglomerate, 
neglecting the contributions due to other possible excitations. Various curves 
correspond to different values of the applied bias, namely different $\omega _0$. Note that in the figure the frequency is rescaled with respect to $\omega_{0}$.

For $\omega_{0}>\omega_{\mathrm{n}}$ (dotted magenta and short-dashed blue curves) 
it is easy to note a pronounced peak in correspondence of $\omega = \omega_0$ due to the single-qp contribution. Despite 
the fact the temperature is very low, thermal effects leads to a smoothening of the peak.  A second contribution, less pronounced, 
is observable at $\omega = 2\omega _0 $,  signature of the presence of the $2$-agglomerate that has a less divergent 
power-law behavior. This secondary peaks becomes more and more evident by increasing $\omega_{0}$. 

For bias voltages such that $\omega_{0}<\omega_{\mathrm{n}}$ (Long-dashed green and solid red curves) the neutral mode contribution increases the 
scaling dimension turning the single-qp peak into an extremely broad dip that completely 
hides the $2$-agglomerate contribution. This is a clear demonstration of the fact that the ideal 
condition to observe the peak patter for the $\nu=2/5$ case is $\omega _0 > \omega _{\mathrm{n}}$. A remark is 
in order concerning the role of the temperature. Indeed, the smoothening induced by thermal effects could 
completely wash away the observed structure under the condition $T>\omega_{\mathrm{n}}$, therefore very low temperature are needed to see this structure.
 
\section{Conclusions}
\label{conclusion}
In this paper we analyzed the finite frequency noise for  the FQH 
state at $\nu = 2/5$.  We considered the presence of the two more relevant excitations: the
single-qp, with charge $e/5$, and the $2$-agglomerate, with charge
$2 e/5$. The finite frequency noise has the unique possibility to resolve \emph{spectroscopically} the
contributions of the different excitations looking at different Josephson resonances. 
We showed that the peak associated to the $2$-agglomerate is more evident at Josephson frequency higher 
than the neutral mode cut-off, where the tail of the single-qp contribution decreases faster. 
We also commented on the evolution of the height of the single-qp peak as a function of temperature, in order to determine the optimal condition for the visibility of the considered peak pattern. 

\section*{Acknowledgements}
We thank M. Carrega, T. Martin and F. Portier for useful discussions. We acknowledge the support of the EU-FP7 via ITN-2008-234970 NANOCTM.

\end{document}